\documentclass[twocolumn,showpacs,preprintnumbers,amsmath,amssymb]{revtex4}

\usepackage{graphicx}
\usepackage{dcolumn}
\usepackage{bm}
\usepackage{here}

\usepackage{amsmath}

\begin{document}
\preprint{}
\title{Recoil-free spectroscopy of neutral Sr atoms in the Lamb-Dicke regime}

\author{Tetsuya Ido${}^{1}$\footnote[1]{Current address: JILA, University of Colorado, CO 80309-0440.} and Hidetoshi Katori${}^{1,2}$\footnote[2]{Author to whom correspondence should be addressed.\\ E-mail address: katori@amo.t.u-tokyo.ac.jp}}
\affiliation{
${}^1$Cooperative Excitation Project, ERATO, Japan Science and Technology Corporation, 4-1-8 Hon-cho, Kawaguchi, Saitama 332-0012, Japan \\ 
${}^2$Engineering Research Institute, University of Tokyo, Bunkyo-ku, Tokyo 
113-8656, Japan.
}

\date{\today}
\begin{abstract}
We have demonstrated a recoil-free spectroscopy on the ${}^1S_0-{}^3P_1$ transition of strontium atoms confined in a one-dimensional optical lattice. 
By investigating the wavelength and polarization dependence of the ac~Stark shift acting on the  ${}^1S_0$ and ${}^3P_1(m_J=0)$ states, we determined the {\it magic wavelength} where the Stark shifts for both states coincide.
The Lamb-Dicke confinement provided by this Stark-free optical lattice enabled the measurement of the atomic spectrum free from Doppler as well as recoil shifts. 
\end{abstract}

\maketitle

Recent dramatic advances in optical metrology have made it possible to directly link two optical frequencies with an uncertainty below $10^{-18}$ \cite{Stenger} and to coherently divide an optical frequency down to a radio frequency defined by the SI second \cite{H1S2S}.
Owing to these techniques, a stringent comparison of the stability and accuracy among optical clocks becomes feasible \cite{Diddams} and leads to an improved definition of time and a test of the time variation of a fundamental constant \cite{Prestage}. 
Up to now, two sorts of absorbers, single ions in the Lamb-Dicke regime (LDR) \cite{Dicke} and neutral atoms in free space, have been extensively studied for optical clocks.
Tightly confined single ions, enabling a long interaction time and the Doppler- as well as the recoil-free absorption \cite{Bergquist87}, have, so far, led to the narrowest optical spectrum \cite{Hg}. 
However, the stability was severely limited by the quantum projection noise (QPN) \cite{QPN} of the single absorber. The ensemble of neutral atoms, in contrast, provides a far better QPN limit. The accuracy of the measurement is known to be affected by the atomic motion, as residual Doppler shifts are introduced through the imperfect wave front of the probe beam \cite{PTB2002,wavefront}.

These two approaches could merge if an ensemble of neutral atoms were separately prepared in the LDR. An optical lattice made by the interference pattern of a light field \cite{Lattice} confines atoms in a small volume, which satisfies the Lamb-Dicke condition \cite{Dicke}. However, an impediment for the precision measurement is the complete control over the perturbations caused by this container, since the Stark shift potential strongly depends on the electronic state, which is not the case for ion-trapping potential.

The transition frequency between the states, $|g\rangle$ and $|e\rangle$, subjected to the ac~Stark shift of the trap laser with an electric field of $\vec{E}(\omega_L, \hat{\epsilon})$, can be written as,
\begin{eqnarray}
\omega_{\rm obs}&=&(\omega_e-\omega_g) \nonumber \\
&-&\Delta\alpha(\omega_L,\hat{\epsilon})|\vec{E}(\omega_L,\hat{\epsilon})|^2/4\hbar +O(\vec{E}^4).
\label{shift}
\end{eqnarray}
Here, $\omega_e-\omega_g (=\omega_0)$ is the unperturbed atomic resonance frequency, and \[ \Delta\alpha(\omega_L,\hat{\epsilon})= \alpha_e(\omega_L,\hat{\epsilon})-\alpha_g(\omega_L,\hat{\epsilon}) \]
denotes the differential dipole-polarizability.
If the polarizabilities $\alpha_e$ and $\alpha_g$ coincide at a specific trapping laser frequency $\omega_L$ and polarization $\hat{\epsilon}$, the second term of Eq.~(1) vanishes.
Therefore, the unperturbed atomic transition frequency, independent of the trapping laser intensity of $I_L\propto|\vec{E}|^2$, can be observed \cite{JPSJ,Scotland}.
This Stark shift cancellation technique was first demonstrated by an improved atom loading into an optical trap for ultracold strontium atoms \cite{JPSJ}.
Similar ideas of tailoring ac~Stark shift potentials have been employed in the Raman sideband cooling \cite{Jessen1998} and fine spectroscopy \cite{Davidson} of alkali atoms in the lower-lying hyperfine states.

In this Letter, we report the recoil-free spectroscopy of strontium atoms confined in a one-dimensional optical lattice by applying the technique to the $(5s^2){}^1S_0-(5s5p){}^3P_1$ ``clock" transition. 
By investigating the wavelength as well as the polarization dependence of the differential dipole-polarizability, we have determined the {\it magic wavelength}, where the Stark shift on the clock transition vanishes, {\it i.e.}, $\Delta\alpha=0$.

In order to discuss the polarization dependence of the Stark shift and the influence of Raman coherences between the $m,\ m'$ magnetic sublevels of the $|a\rangle$ state, we introduced the light-shift Hamiltonian of $V_{m'm}$. We assumed electric-dipole couplings to electronic states of $|b\rangle$ by a far-detuned radiation with intensity $I_L$, 
\begin{eqnarray}
V_{m'm}& =& m\mu B \delta_{m'm} \nonumber \\
&- &3\pi c^2 I_L \sum_{b}\frac{\gamma_{ba}\Lambda_{m'm} (b, \hat{\epsilon})}{\omega_{ba}^2(\omega_{ba}^2-\omega_{L}^2)}.
\label{lightshift}
\end{eqnarray}
The first term of Eq.~(\ref{lightshift}) denotes the Zeeman energy in the magnetic field of $B$, and the second term denotes the light shifts, where $\gamma_{ba}$ and $\omega_{ba}$ are the natural linewidth and the transition frequency for the $|b\rangle\rightarrow |a\rangle$ transition, respectively.
The summation was carried out over the excited states, as described in our previous work \cite{JPSJ}.
$\Lambda ({b},\hat{\epsilon})$ denotes the coupling matrix element with the normalized dipole moment of $\hat{\bf d}_{ba}$ \cite{CCT}.
\begin{equation}
\Lambda_{m'm}(b, \hat{\epsilon})=\sum_{m_b}\langle m'| \hat{\epsilon} \cdot \hat{\bf d}_{ba} | m_b\rangle \langle m_b | \hat{\epsilon^*} \cdot \hat{\bf d}_{ba}^{\dagger}| m \rangle, \nonumber
\end{equation}
in which the summation runs over the magnetic sublevels of $m_{b}$ in the $|b\rangle$  state accessed by the light polarization of $\hat{\epsilon}$.
For the $a={}^1S_0$ state,  $\Lambda (b, \hat{\epsilon})$ does not show any polarization dependence because of the isotropy of the state.
However, for the upper state of the ``clock transition" $a={}^3P_1 (m_J=0)$, the coupling strength of $\Lambda(b, \hat{\epsilon})$ strongly depends on the light polarization.
This is illustrated in Fig.~1(a) by the thickness of the arrows: taking the ${}^3S_1$ state as the $|b\rangle$ state, this state is not coupled by the $\pi$ polarized light but by the $\sigma^\pm$ polarized light.
In this way, the Stark shift for the $^3P_1$ state is critically affected by the applied light polarization.

\begin{figure}
\includegraphics[width=\linewidth]{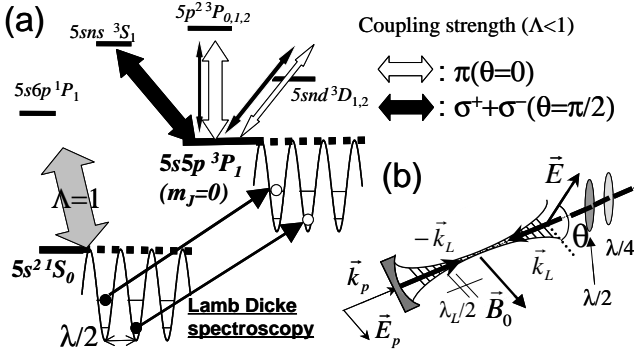}
\caption{(a) Level diagram and optical coupling related to the clock transition of ${}^1S_0-{}^3P_1$. The ${}^1S_0$ state is dominantly coupled to ${}^1P_1$, while the ${}^3P_1$ state is coupled to the upper triplet states of ${}^3S, {}^3P^e$, and ${}^3D$. The latter couplings strongly depend on the polarization of the laser field, as depicted by the thickness of arrows.
(b) Schematic diagram of the 1D optical lattice. A bias magnetic field $\vec{B}_0$ was applied parallel to the probe electric field $\vec{E}_p$ with a wave vector  $\vec{k}_p$ parallel to that of the lattice beam $\pm\vec{k}_L$. $\theta$ represents an angle between $\vec{B}_0$ and the electric field $\vec{E}$ of the lattice.
The probe laser was introduced parallel to  $\pm\vec{k}_L$, along which the spatial modulation of the ac Stark shift confines atoms in the LDR, as 
depicted in (a).
}
\end{figure}

Our experimental setup and the procedure to form an optical lattice have been described elsewhere \cite{PRA, Muka}.
A few $\mu$K cold strontium atoms were magneto-optically cooled and trapped on the narrow $^1S_0-{}^3P_1$ transition \cite{SrPRL}.
During the final cooling and trapping period of 30 ms, the atoms were loaded into a one-dimensional optical lattice.
A frequency-stabilized Ti-Sapphire laser (Coherent, MBR-110) was coupled into a polarization-maintaining optical fiber and focused into the atom cloud by an objective lens.
A quarter-wave and a half-wave plate were used to compensate the birefringence of the optical fiber and obtain a linearly polarized light with an arbitrary angle, as shown in Fig.~1(b).
This trap beam, with a typical power of 500 mW, was then retro-reflected by a concave mirror with a radius of 25 cm to form a standing wave, where the $1/e$ radius of the beam waist was $\approx 23\ \mu{\rm m}$ at $\lambda_L=$ 830 nm.

With these parameters, the axial confinement frequency in the $^1S_0$ state was measured to be $\nu_S=90\ {\rm kHz}$ by observing the first heating sideband, as discussed later. This vibrational frequency suggested a peak laser intensity of $I_{\rm peak}\approx 47 \ {\rm kW/cm}^2$, a depth of $U=k_{\rm B}\times 30\ \mu{\rm K}$ or $h\times610\ {\rm kHz} $, and a radial trap frequency of $\nu_r=460\ {\rm Hz}$, respectively.  
Figure 1(b) shows a configuration for the spectroscopy. The confined atoms were excited by a probe laser with a wave vector of $\vec{k}_p$, which was introduced through the dichroic concave mirror and superimposed on the lattice beam with $\pm \vec{k}_L$.
The axial confinement of $\nu_S$ gave the Lamb-Dicke parameter of $\eta = \sqrt{h/2M\nu_S}/\lambda_{\rm CLK}= 0.24$, where $M$ is the atomic mass and $\lambda_{\rm CLK}= 689\ {\rm nm}$,  the wavelength of the $^1S_0-{}^3P_1$ clock transition.
A bias magnetic field of $|\vec{B}_0|=0.5 \ {\rm G}$ was applied parallel to the probe electric field of $\vec{E}_p$ to define the quantization axis, which split the $m=\pm1$ sublevels of the $^3P_1$ state by $\pm \mu |\vec{B}_0|/h\approx \pm 1\ {\rm MHz}$.
The direction of the electric field $\vec{E}$ of the lattice laser was defined by an angle $\theta$ with respect to the quantization axis defined by $\vec{B}_0$.

\begin{figure}
\includegraphics[width=0.9\linewidth]{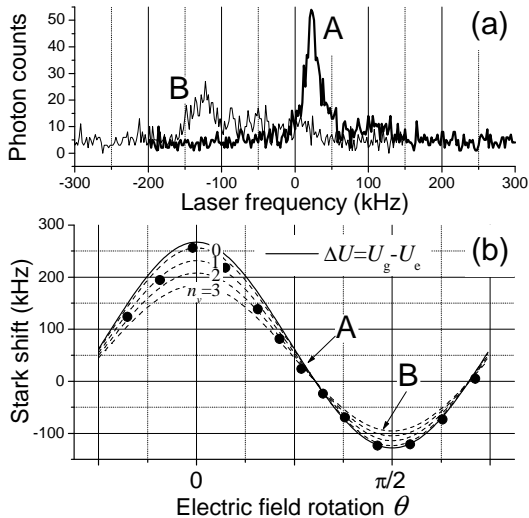}
\caption{(a) Fluorescence spectra of Sr atoms confined in a 1D optical lattice for different polarization rotation $\theta$ at $\lambda_L=830\ {\rm nm}$. 
At $\theta\approx \pi/4$ (A), the narrower spectrum was observed as the Stark shifts on the clock transition nearly canceled out, whereas non-zero differential Stark shift introduced a frequency shift and broadening for $\theta\approx\ \pi/2$ (B). 
(b) The change of the Stark shift as the rotation $\theta$ of the lattice electric field. 
Since the light shift of the ${}^3P_1 (m_J=0)$ depends on the lattice laser polarization, the frequency varies sinusoidally depending on the components of $\pi$ and the $\sigma^++\sigma^-$ intensity.
The dashed lines show the differential Stark shifts for atoms in the $n_v=0,1,2$ vibrational states, see text. 
}
\end{figure}

Figure~2(a) shows laser-induced fluorescence spectra for different electric field rotations with $\theta\approx\pi/4$ (A) and $\theta\approx\pi/2$ (B) at $\lambda_L=830\ {\rm nm}$.
The probe intensity was $I_p\approx 3I_0$ with $I_{0}= 3\ \mu{\rm W/cm^2}$, the saturation intensity of the clock transition. 
The main peaks correspond to the $|^1S_0,n_v\rangle\rightarrow|^3P_1,n_v\rangle$ transition, where $n_v(=0,1,2\cdots)$ represent the vibrational state of atoms in the Stark shift potential.
The frequency difference of the axial-confinement $\delta \nu=\nu_P-\nu_S$ for the $^3P_1$ and ${}^1S_0$ states caused a vibrational-state-dependent frequency shift of $n_v \delta \nu $: The spectrum (B) suffered an inhomogeneous broadening because of the measured thermal distribution of $\langle n\rangle \approx3.7$.
In contrast, a simple picture of sideband cooling can be applied for the case of A, where the vibrational frequency was nearly degenerate.
A small peak at 100 kHz corresponded to the excitation of the $\Delta n_v=+1$ heating sideband, while the $\Delta n_v=-1$ excitation (cooling sideband) was not visible as the atoms were quickly sideband-cooled to the vibrational ground state \cite{Wineland89}. 

The change of the resonance frequency is summarized in Fig.~2(b) as a function of the lattice laser polarization of $\theta$.
Since the ${}^1S_0$ state has no polarization dependence, the frequency change is attributed to the ac~Stark shift of the ${}^3P_1(m_J=0)$ state. 
The solid curve in Fig.~2(b) shows the differential Stark shift obtained from Eq.~(\ref{lightshift}), and the dashed lines, the corresponding Stark shifts for the $\Delta n_v=0$ vibrational transitions with $n_v=0,1,2$.
The vertical spread of the curves explains the broadening of the spectrum, as discussed previously. The observed frequency shift should be smaller than the $n_v=0$ differential Stark shift because of the thermal distribution of typically $\langle n\rangle \approx3.7$.
Since the applied Zeeman shift of $\approx 1$ MHz was strong enough to suppress the Raman coherence among Zeeman sublevels, the polarization dependence mainly accounted for the intensity of the $\pi$ polarized light ($\theta=0$) and the linear combination of $\sigma^+ +\sigma^-$ ($\theta=\pi/2$), resulting in a sinusoidal change. 
As the deviation from the theoretical curve was within 5 \%, in the following analysis to determine the light shifts for $\theta=0$ and $\theta=\pi/2$, we approximated the fitting curve by a sinusoidal function.

\begin{figure}
\includegraphics[width=0.9\linewidth]{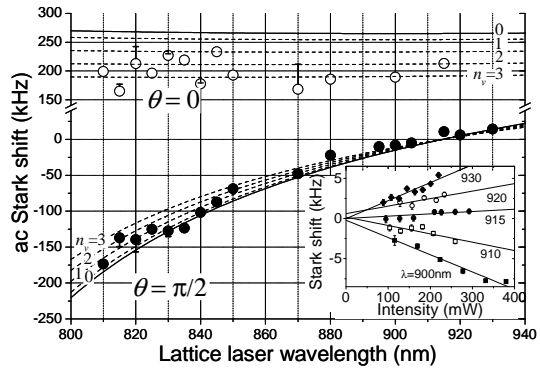}
\caption{Differential Stark shift for the $^1S_0-{}^3P_1(m=0)$ transition.
The shift depends on the trap laser wavelength $\lambda_L$, the polarization rotation $\theta$, and the vibrational state $n_v$, indicated by dashed lines. 
The inset shows the intensity dependence of the shifts measured for $\theta=\pi/2$ around the {\it  magic wavelength} of $\lambda_L=915\ {\rm nm}$. By linearly extrapolating to zero laser power, the atomic resonance frequency was obtained.}
\end{figure}

Figure 3 shows the Stark shifts for the $\theta =0$ and $\pi /2$ in the range of $\lambda_L=810 - 930$ nm, where the Stark shift was normalized with the laser intensity of $I_{\rm eff}\approx 47 \ {\rm kW/cm}^2$.
We note that, in the wavelength range of $\lambda_L=(690)-915 \ {\rm nm}$, in which the Stark shifts for the $\theta=0$ and $\pi/2$ have an opposite sign, the differential polarizability $\Delta\alpha$ may vanish for a specific light polarization angle of $\theta$, as demonstrated in Fig.~2(a). 
The observed efficient loading of atoms from the magneto-optical trap into the optical trap at $\lambda_L\sim 800\ {\rm nm}$ \cite{JPSJ,PRA} can be attributed to this angle-tuning mechanism.

In order to accurately determine the unperturbed atomic resonance frequency $\omega_0$ free from the Stark shifts, we measured the power dependence of the transition frequencies at several $\lambda_L$ for $\theta=\pi/2$, as shown in the inset of Fig.~3.
By linearly extrapolating $I_L\rightarrow 0$, the measurements at different wavelengths converged within 1 kHz, which was limited by the center frequency reproducibility of our laser system \cite{Li}.
We took this value as a reference for the atomic resonance frequency and determined the {\it magic wavelength} for $\theta=\pi/2$ to be $\lambda_L=914\pm1\ {\rm nm}$.
The inset also indicated that a higher-order Stark shift due to hyperpolarizability did not make a significant contribution in this intensity range, thus justifying the normalization of the Stark shifts by the trap laser intensity.

We used this {\it magic wavelength} to determine the uncertain transition moments in the calculation of the Stark shifts.
Among the transition strengths listed in Ref. \cite{Acoeff}, the $(5s6s){}^3S \rightarrow (5s5p){}^3P$ transition rate showed a rather large discrepancy of more than $30\ \%$.
Therefore, we took this rate as a fitting parameter to locate the {\it magic wavelength} at 914 nm. 
The estimated transition rate of $A=8.5\times10^{7}/{\rm s}$ was within the reported values in the range of $ 6.7-9.2 \times10^{7}/{\rm s}$ \cite{Acoeff}.
Using the value, we calculated the differential Stark shifts shown by the solid lines in Fig.~3.
The difference between the data points and theoretical curve is attributed to 1) the imperfect lattice laser polarization, as indicated by error bars in the figure, 2) the uncertainty in determining the center frequency of the spectrum due to the vibrational state spread, as indicated by dashed lines, and 3) the day-to-day change of the beam overlap of the lattice standing wave and resultant intensity fluctuation.
We note, however, the determination of the magic point was rather accurate, since the point is independent of the laser intensity fluctuation and 
occupied vibrational states of confined atoms.

\begin{figure}
\includegraphics[width=0.9\linewidth]{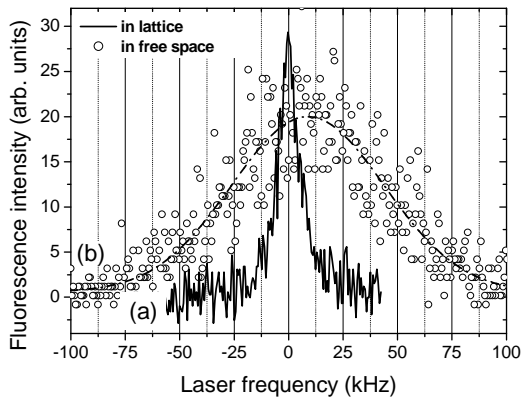}
\caption{Laser-induced fluorescence of atoms confined in a 1D optical lattice (a) and in a free space (b). The dashed line shows a Gaussian fit to the data points (b). The confinement suppressed the Doppler width of 83 kHz and gave a narrow Lorentzian linewidth of 11 kHz, which was limited by the saturation broadening. 
A slight blue shift of the center frequency in (b) is caused by the photon recoil shift; see text.}
\end{figure}

Figure 4 demonstrates a recoil-free spectrum for $\theta = \pi/2$ at $\lambda_L= 915\ {\rm nm}$, 
in which the Stark-free condition is nearly satisfied.
The atom number was decreased to $\sim 10^4$, or $\sim 10$ atoms in each lattice site, to moderate the attenuation of the probe beam and the collision shifts.
We interrogated the atoms for $\tau_p=2\ {\rm ms}$ in each measurement and averaged 32 scans. 
The measured linewidth of 11 kHz (FWHM) was in agreement with the saturation broadened linewidth for the applied probe intensity of $I_p\approx I_0$.

In order to demonstrate the modification of the atomic spectrum by the LDR, we measured the laser-induced fluorescence of free atoms.
Figure 4(b) shows the fluorescence for a probe duration of $\tau_p=1.0\  {\rm ms}$, applied at 1 ms after turning off the optical lattice.
The Doppler width of 83 kHz (FWHM) suggested the atom temperature of $6\ \mu {\rm K}$ in the lattice. 
The slight blue shift of $\delta f=9\ {\rm kHz}$ was attributed to the photon recoil shift as well as the recoil heating of atoms, since more than a single photon was scattered during the probe period of $\tau_p$.
By reducing the duration $\tau_p$ and extrapolating $\tau_p\rightarrow0$, we obtained $\delta f(\tau_p\rightarrow 0)=5.2\pm0.6\ {\rm kHz}$, in agreement with  the recoil shift of  $h / (2M\lambda_{\rm CLK}^2) = 4.8\ {\rm kHz}$.  

In summary, we have demonstrated the recoil-free spectroscopy of neutral $^{88}{\rm Sr}$ atoms confined in the Lamb-Dicke regime provided by a 1D optical lattice.
By adjusting the Stark shifts for the probed upper and lower states, we canceled out the perturbation of the lattice potential on the atom spectrum, thus realizing a system similar to ion-trapping experiments \cite{Bergquist87}. 
The application of this scheme to a frequency standard is promising because simultaneous preparation of $N$ neutral atoms in a 3D optical lattice improves the quantum projection noise by a factor of $\sqrt{N}$ compared to the single-ion-based standard \cite{Hg}.
The frequency accuracy of the present scheme may finally be limited by the purity of the lattice-laser polarization. The use of the ${}^1S_0-{}^3P_0$ transition of ${}^{87}{\rm Sr}$ dramatically reduces the polarization dependence \cite{Scotland}, which may provide an alternative for the optical-frequency standard.

We thank M. Kuwata-Gonokami for valuable comments and support and Y. Li and T. Mukaiyama for their participation in discussions and assistance with the experiments.


\begin{references}
\bibitem{Stenger}
J. Stenger, H. Schnatz, C. Tamm, and H.R. Telle, Phys. Rev. Lett. {\bf 88}, 073601 (2002).
\bibitem{H1S2S}
M. Niering, R. Holzwarth, J. Reichert, P. Pokasov, T. Udem, M. Weitz, T. W. H\"{a}nsch, P. Lemonde, G. Santarelli, M. Abgrall,  P. Laurent, C. Salomon, and A. Clairon,  Phys. Rev. Lett. {\bf 84} 5496 (2000). 
\bibitem{Diddams} S. A. Diddams, T. Udem, J. C. Bergquist, E. A. Curtis, R. E. Drullinger, L. Hollberg, W. M. Itano, W. D. Lee, C. W. Oates, K. R. Vogel, and D. J. Wineland, Science {\bf 293}, 825 (2001).
\bibitem{Prestage}
J. D. Prestage, R. L. Tjoelker, and L. Maleki, Phys. Rev. Lett. {\bf 74} 3511 (1995).
\bibitem{Dicke}
R. H. Dicke, Phys. Rev. {\bf 89} 472 (1953).
\bibitem{Bergquist87}
J. C. Bergquist, W. M. Itano, and D. J. Wineland, Phys. Rev. A {\bf 36} R428 (1987). 
\bibitem{Hg} R. J. Rafac, B. C. Young, J. A. Beall, W. M. Itano, D. J. Wineland, and J. C. Bergquist, 
Phys. Rev. Lett. {\bf 85}, 2462 (2000).
\bibitem{QPN}
W. M. Itano, J. C. Bergquist, J. J. Bollinger, J. M. Gilligan, D. J. Heinzen, F. L. Moore, M. G. Raizen, and D. J. Wineland, Phys. Rev. A {\bf 47}, 3554 (1993). 
\bibitem{PTB2002} G. Wilpers, T. Binnewies, C. Degenhardt, U. Sterr, J. Helmcke, and F. Riehle, Phys. Rev. Lett. {\bf 89}, 230801 (2002).
\bibitem{wavefront} T. Trebst, T. Binnewies, J. Helmcke, and F. Riehle, IEEE Trans. Instrum. Meas. {\bf IM 50}, 535 (2001). 
\bibitem{Lattice}
P. S. Jessen and I. H. Deutsch, Adv. At. Mol. Opt. Phys. {\bf 37}, 95 (1996).
\bibitem{JPSJ}
H. Katori, T. Ido, and M. K.-Gonokami, J. Phys. Soc. Jpn. {\bf 68}, 2479 (1999).
\bibitem{Scotland}
H. Katori, {\it in the Proceedings of the 6th Symposium on Frequency Standards and Metrology}, (World Scientific Publishing Co., 2002) pp.323-330. 
\bibitem{Jessen1998}
S. E. Hamann, D. L. Haycock, G. Klose, P. H. Pax, I. H. Deutsch, and P. S. Jessen,
Phys. Rev. Lett. {\bf 80}, 4149 (1998);
V. Vuleti\'{c}, C. Chin, A. J. Kerman, and S. Chu, Phys. Rev. Lett. {\bf 81}, 5768 (1998).
\bibitem{Davidson} A. Kaplan, M. F. Andersen, and N. Davidson, \pra {\bf 66}, 045401 (2002).
\bibitem{CCT}
C. Cohen-Tannoudji, "Fundamental Systems in Quantum Optics" in Les Houches, edited by J. Dalibard, J. M. Raimond, and J. Zinn-Justin (North-Holland, 1992). 
\bibitem{PRA}
T. Ido, Y. Isoya, and H. Katori, Phys. Rev. A {\bf 61}, 061403(R) (2000).
\bibitem{Muka}
T. Mukaiyama, H. Katori, T. Ido, Y. Li, and M. K.-Gonokami, to be published in Phys. Rev. Lett.
\bibitem{SrPRL}
H. Katori, T. Ido, Y. Isoya, and M. Kuwata-Gonokami, Phys. Rev. Lett. {\bf 82}, 1116 (1999).
\bibitem{Wineland89} F. Diedrich, J. C. Bergquist, W. M. Itano, and D. J. Wineland, Phys. Rev. Lett. {\bf 62}, 403 (1989).
\bibitem{Li} Y. Li, T. Ido, T. Eichler, and H. Katori, to be submitted.
\bibitem{Acoeff}
H. G. C. Werij, C. H. Greene, C. E. Theodosiou, and A. Gallagher, \pra {\bf 46}, 1248 (1992). 
\end{references}
\end{document}